# Temperature Variability and Natural Disasters

23 September 2024

Aatishya Mohanty [a], Nattavudh Powdthavee [b,d], CK Tang [b], Andrew J. Oswald[c,d,e*]

[a] *University of Aberdeen, UK, AB24 3QY.*
[b] *Nanyang Technological University, Singapore, 639798.*
[c] *University of Warwick, UK, CV4 7AL.*
[d] *IZA Institute, Bonn 53113, Germany.*
[e] *CAGE Research Centre, UK, CV4 7AL.*

[*]To whom any correspondence can be addressed: Telephone: +44 1437729452. Email: andrew.oswald@warwick.ac.uk. Full address: University of Warwick, Economics Department, Coventry CV4 7AL, United Kingdom.

ACKNOWLEDGEMENTS: A special vote of thanks for his comments goes to the climate scientist Professor Sir Brian Hoskins. We also wish to thank Menghan Yuan, Amanda Goodall, and David Stainforth.
WORD COUNT: 8000 words approx. (excluding tables and the Appendix with supplemental material).
KEY WORDS: temperature; standard deviation; climate change; mental health; wellbeing; environment; green.
AUTHOR CONTRIBUTIONS: All authors designed the analysis and agreed the conclusions.
DATA AVAILABILITY: The data sets are publicly available.
AI: AI was not used in any of this work or the manuscript.
FUNDING STATEMENT: No explicit funding source.
CONFLICT OF INTEREST: The authors declare none.
ETHICS APPROVAL: Not applicable, although approval was gained by the original data-set collectors.
PATIENT CONSENT STATEMENT: Not applicable.
PERMISSION TO REPRODUCE FROM OTHER SOURCES: Not applicable.
OTHER VERSIONS OF THIS PAPER: None.

## Abstract

This paper studies natural disasters and the psychological costs of climate change. It presents what we believe to be the first evidence that higher *temperature variability* and not a higher level of temperature is what predicts natural disasters. This conclusion holds whether or not we control for the (incorrectly signed) impact of temperature. The analysis draws upon long-differences regression equations using GDIS data from 1960-2018 for 176 countries and the contiguous states of the USA. Results are checked on FEMA data. Wellbeing impact losses are calculated. To our knowledge, the paper's results are unknown to natural and social scientists.

**KEY WORDS**: Global warming; temperature standard deviation; human wellbeing; happiness; disasters; BRFSS; WVS.
**JEL**: Q54; I31

*Rising temperatures are fueling environmental degradation, natural disasters, weather extremes, food and water insecurity, economic disruption, conflict, and terrorism.* www.un.org/en/un75/climate-crisis-race-we-can-win

*With increasing global surface temperatures the possibility of more droughts and increased intensity of storms …[is]…likely.* www.usgs.gov

*Temperature, in particular, exerts remarkable influence over human systems at many social scales; heat induces mortality, has lasting impact on fetuses and infants, and incites aggression and violence while lowering human productivity.* Carleton and Hsiang, Science, 2016.

## Introduction

This paper examines the patterns in sixty years of data on climate change. The cornerstone of the analysis is the foundational importance of the variability of the world's temperature. Later tables provide what we believe to be the first formal econometric evidence that temperature variability, and not the level of temperature, is predictive of natural disasters. A non-technical reader can find a flavor of the paper's main conclusions in the scatter plots of Figures 1-6.

The paper differs from a long tradition in public and informal discussion. Famously, the focus has been upon the idea that a higher level of temperature can cause greater numbers of natural disasters.[1] We argue differently. Our key finding -- on the much stronger role of variability -- holds true in a range of settings across countries and regions, is robust to different specifications (natural log, inverse hyperbolic sine, etc), is robust to different disaster data sets, and is not driven simply by the occurrence of extreme temperatures at the upper end [2] of the distribution. The various data sources used here cover the period 1960-2018 for 2355 regions in 176 countries and separately for the 49 contiguous states within the United States of America.

The concept of climate -- as distinct from weather -- is sometimes empirically approximated in research by taking at least an average over 30 years (Pachauri et al, 2014).

[1] These statements should be kept in perspective. Climate scientists are aware, of course, of the wider nature of the altering climate. "Global warming involves changes not only in the mean atmospheric temperature, but also in its variability and extremes.": Tamarin-Brodsky et al. (2020).

[2] If low temperatures stay approximately the same and high temperatures increase then, simply arithmetically, the standard deviation of temperature, for example, will rise. But that is not the reason for this paper's finding.

Although Hsiang (2016) rightly notes that this definition has an arbitrary aspect, we follow it, for sheer data reasons, and adopt a 30-year averaging principle as part of a 'long differences' analysis. However, the flavor of our findings does not depend on this exact way to calculate a climate, rather than weather, variable[3].

The scientific background is familiar. Greenhouse-gas emissions in the world are rising. "With every additional increment of global warming, changes in extremes continue to become larger… Compound heatwaves and droughts are projected to become more frequent … intensification of tropical cyclones and/or extratropical storms… and increases in aridity and fire weather" (IPCC Synthesis Report 2023). Economists, as well as climatologists, continue to alert us (Schmalensee et al. 1998, Weitzman 2009, Stern 2008, 2018, 2022, Nordhaus 2019). It is known that attendant natural disasters, which many scientists believe will become more common with climate change, are costly in human and animal lives, health, and economic consequences (Kahn 2005, Luechinger and Raschky 2009, Noy 2009, Rehdanz et al. 2015, Sekulova and van den Bergh 2016, Ahmadiani and Ferreira (2021), Noy et al. 2021, Noy and Strobl 2023, Cavallo et al. 2013, Bentzen 2019, Coronese et al. 2019, Boustan et al. 2020, Carvalho et al. 2021, Castle et al. 2022, Stein and Weisser 2022, Bertinelli et al. 2023, Frijters et al. 2020, Knight 2023, Noy et al. 2023). It should be noted, however, that Buggle and Durante (2021) argue that climate risk may also lead to greater trust and cooperation among people, and Noy and Strobl (2023) for the possibility that innovation might be one beneficial consequence of disasters.

On the balance of the modern evidence, in the view of some researchers, a feature of climate change is that temperature variability is likely to become more prevalent in the future

[3] We explored 10-year averaging instead, for example. But the principal approach we use is to calculate average temperature in each region within a country for each year. (In some of our checks we also used areas of the world considerably smaller than states of the USA.) Then we take the standard deviation SD across the first block of thirty years, and take the difference compared to the SD across the next block of thirty years. So these are 'long differences' in SD.

(particularly in some parts of the world: Bathiany et al. 2018, Rodgers et al. 2021, Shi et al. 2024). Technical research on that question is continuing. See especially Tamarin-Brodsky et al. (2020). Earlier estimates by Huntingford et al. (2013), however, concluded that the average level of climate variability had stayed approximately constant through time.

The objective of the current paper is to aim for an understanding of climate itself rather than an understanding of weather shocks (a single extreme temperature day or other exceptional shock in time t). That is an inherently difficult task within data sets covering comparatively small numbers of years. Deliberately, therefore, the later analysis will not put all its emphasis upon one-year movements[4] but instead, very substantially, upon averages over a fairly long period. We will be especially concerned, like Robeson (2002), with any potential overlap between mean and standard deviation of air temperature -- and their respective implications for global warming and human outcomes. We do not study skewness (Tamarin-Brodsky et al. 2019 point out that climate researchers have paid relatively little attention to higher moments of the temperature distribution).

Later analysis will concentrate on (two kinds of) temperature variables. It will not consider other potential climatic measures[5]. This is partly for data reasons and partly for strategic reasons, to help make the project manageable, by narrowing the focus of the study to a reduced-form examination of issues of global warming.

## Background

The paper blends data on climate, natural disasters, and measures of human wellbeing.

---

[4] Researchers can, we think, legitimately proceed in different ways in the study of climate change. Hsiang (2016) is arguably the best exposition of the statistical issues and includes an interesting discussion of the envelope theorem. Our treatment of the problem in this paper can be seen, under his three-way classification of methods, as a version of long-differences.

[5] We effectively view the two temperature variables (level and standard deviation) as exogenous over the decades on which we have data. From the start of the project we wished to avoid potential issues of so-called p-hacking. We have not attempted to include in the regressions other independent climate variables, such as precipitation or wind speed, and are aware that climatic conditions interact through a complex and not fully understood system. Our focus is temperature.

On the latter, we follow partly in the footsteps of early work such as Maddison and Rehdanz (2011). Weather kinds of variables and especially temporary spikes in temperature are known to affect life satisfaction scores, mental health, and other reported human feelings (Feddersen et al. 2016, Von Möllendorff and Hirschfeld 2016, Baylis et al. 2018, Obradovicha et al. 2018, Pailler and Tsaneva 2018, Mullins and White 2019, Baylis 2020, Frijters et al. 2020, Lawrance et al. 2022, Gunby and Coupe (2023), Lignier 2023, Thompson et al. 2023, although some estimated effects are minor: eg. Hailemariam et al. (2023). They have also been shown to have important and statistically significant links to suicides and violence (Burke et al. 2018, Harari and La Ferrara 2018, Baysan et al. 2019) and to economic variables such as output and productivity (Deschênes and Greenstone (2007), Hsiang 2010, Dell et al. 2014, Burke et al. 2015, Calel et al. 2020).

Stern (2008), Tol (2009), and Hsiang and Kopp (2018) are valuable background reviews for economists. Stainforth (2023) is a recent exposition of the scientific difficulties involved in knowing how evolving alterations in climate will affect the world.

Frumkin et al. (2008) explains the seriousness of the public health problems created by global warming. Berlemann (2016) and Berlemann and Eurich (2021), following in the original footsteps of Kimball et al. (2006), demonstrate, using Gallup data over more than a decade, that hurricane risk leads to reduced levels of life satisfaction across the areas (at the zip-code level) of the USA.

Later estimates in this paper, of the influence of temperature variability upon human lives, do link, to some degree, either directly or indirectly to the spirit of some previous research. Xue et al. (2019) examines data on approximately 20,000 Chinese citizens at two points in time. The authors find that people's mental health declined when air quality, proxied by PM 2.5 particulates, became worse, but also when there was a rise in the variability of temperature (measured by the standard deviation within one year). Xue et al. cannot detect any effect from the mean temperature level. Schär et al. (2004) is unusual in its focus, like that of the present study, upon the variability of temperature. It concludes: “our results demonstrate that the European summer climate might experience a pronounced increase in year-to-year variability in response to greenhouse-gas forcing. Such an increase in variability might be able to explain the unusual European summer 2003, and would strongly affect the incidence of heatwaves and droughts in the future”. Mumenthaler et al. (2021) shows that the number of tweets increases when temperature’s variability increases.

Particular mention should be made of the work of the economist Michael Donadelli, whose prior research, although of a different kind from that here, also in some cases uses the standard deviation of temperature over a decade (we were initially unaware of this work within macroeconomics). Donadelli et al. (2017) examines shocks in a VAR model and studies how a one standard deviation change to the level of temperature tends to reduce growth. Donadelli and colleagues (2021) conclude, using a long series of UK annual data, that there are welfare losses from intra-annual temperature variability, but that there is no consistent evidence on the sign of the effect from temperature variability and economic productivity. Donadelli et al. (2022) is an analysis of cross-country data. It generates mixed results, depending on the continent studied, but broadly finds that summer inter-annual temperature variability, judged by a 10-year standard deviation measure, leads to a drop in productivity in both Europe and North America.

## Data

For this paper the data on climate are taken principally, following many other researchers in the world, from the Climatic Research Unit (CRU) Time-Series database archived at the University of East Anglia in England. The CRU dataset provides monthly gridded weather data over the period 1901-2019. It contains several climate variables for grid cells of the size 0.5 x 0.5 degree (approximately 55.55km by 55.55km). Using this dataset, we construct temperature variables for the sub-national regions corresponding to the locations where the respondents from the surveys are located. We assign temperature measures to the different regions based on the centroid of the grids that are within the respective region boundaries, and compute an aggregate temperature measure for each region by averaging the temperature for all grids by region over each year.[6]

Two outcome measures are of particular interest in the current paper. One is the extent of natural disasters (floods, storms, droughts, hurricanes) and the other is human psychological wellbeing (life satisfaction, happiness).

Information on natural disasters comes principally from the Geocoded Disasters (GDIS) dataset, which is held at the Centre for Research on the Epidemiology of Disasters (CRED) Emergency Events Database (EM-DAT). This dataset reports the location of 9,924 disasters that occurred worldwide from 1960 - 2018. Natural disasters covered by the database include floods, storms (typhoons, monsoons etc.), earthquakes, landslides, droughts, volcanic activity, and extreme temperatures. Based on the latitude and longitude for each recorded disaster, we match the events to the different sub-national regions defined within the World Values Survey (WVS) data set.

A second data set on natural disasters is available, and is used later, for the United States. That information comes from the Federal Emergency Management Agency (FEMA).

The World Values Survey is a cross-country longitudinal survey that measures public attitudes, values, and beliefs on social, economic, political, and cultural subjects.[7] This survey comprises more than 200,000 observations on randomly sample citizens' life satisfaction and happiness, and also on people's personal characteristics and a range of socio-economic outcomes. These are collected at the individual level from 105 countries and spanning 7 waves: 1981-1984, 1989-1993, 1994-1999, 1999-2004, 2005-2009, 2010-2014, and 2017-2020.

[6] For regions that are smaller than the grid (e.g., Hong Kong), we compute the distance to the nearest grid and assume that measured temperature accurately reflect climatic conditions for the area.
[7] Several studies in the literature have relied on the WVS to measure subjective wellbeing and life satisfaction. These includes Inglehart and Klingemann (2000), Maddison and Rehdanz (2011), and Berlemann (2016).

For the later analysis, it is necessary for geo-coded climate data to be accurately matched at the sub-national level to the survey data on individuals sampled in the WVS. We are able to achieve that for sub-national regions within a sample of 93 countries over the period 1990 to 2019. The full list of countries used in the analysis is provided in the supplementary materials. We do the equivalent, over a shorter time span, using data from the Behavioral Risk Factor Surveillance System -- explained below -- on the counties and states of the United States.

One objective is to assess the psychological harm from climate-change phenomena. A WVS question, which has been used by previous researchers, asks respondents to rate their satisfaction with their lives on a scale of 0 to 10, from "dissatisfied" (lowest value on the scale) to "satisfied" (highest value). Likewise, an additional question asks respondents to indicate how happy they feel over the last one year. The responses here range from "not happy at all" (taking a value of 0) to "very happy" (taking a value of 3).

Life satisfaction data on randomly sampled American citizens is available from the Behavioral Risk Factor Surveillance System (BRFSS) database. The BRFSS is a health and behavioral risk survey conducted in the United States by the Centers for Disease Control and Prevention (CDC). It has run annually since 1984. Sample size is approximately 350,000 people per year. The survey gathers data on risk behaviors and on preventive health practices that may have an impact on an adult's health status.

In this paper we use the BRFSS question “In general, how satisfied are you with your life?” to construct a measure of life satisfaction. The answers are on a scale of one to four and can range from “very dissatisfied” (lowest value on the scale) to “very satisfied” (highest value). Our sample here contains slightly less than two million randomly sampled American citizens from 2439 U.S. counties during the period 2005 to 2011.

## Empirical Strategy

For both the international sample and the U.S. sample we estimate disaster equations in so-called long differences. These are regression equations in which the dependent variable is the change over time -- comparing one thirty-year period to another thirty-year period -- in the number of natural disasters in a sub-national geographical area. Equivalently, the independent variables are the change over the same long period in two kinds of temperature variables (we examine mean temperature and the standard deviation of temperature). Hence the format can be thought of as a simple and very particular kind of fixed-effects estimator. Burke and Emerick (2016) and Hsiang (2016) discuss the case for long differences in the study of climate change.

We also study wellbeing equations. We adopt a somewhat similar form of econometric structure as for natural disasters, but with one major difference. In order to estimate impact effects from natural disasters on to citizens, we use conventional annual fixed-effects methods, along with a set of extra controls (for individual respondents' characteristics) that are standard in the empirical literature on wellbeing equations.

For both disasters equations and wellbeing equations, the above specifications imply that the estimated structures are reduced-form ones (as has been true for almost all previous social-science research in the area of climate-change research). That has disadvantages. A variety of questions, about the nature of possible transmission mechanisms, inevitably remain unanswered in reduced-form econometric work. Nevertheless, such an approach seems appropriate at the current juncture in research.

## Main Results

This paper is an inquiry into, in particular, whether the variability of temperature matters independently of the level of temperature.

We begin, in Table 1, with results using total-disasters regression equations. These are estimated in long-difference form and cover 2355 sub-regions across 176 countries for the years between 1960 and 2018. The average number of regions per nation is therefore approximately 13.

Here the long differences are derived in the following way. They are calculated by subtracting the average value of a variable over 1960-1989 from its average value over 1990-2018. As a dependent variable, all three columns of Table 1 use the long-difference in the natural logarithm of total disasters. Each column of Table 1 differs, however, in its independent variables. Those variables are the standard deviation of temperature (labelled SD Temperature), the average temperature (labelled Mean Temperature), and then in the third column a combination of the two.

It can be seen in Table 1 that the level of temperature fails to enter positively, either on its own in column 2 or when the level of temperature is also entered as an independent covariate. Its estimated coefficient has a large standard error. There is no obvious evidence here that the sub-regions of the world are having more disasters as their mean temperature goes up over decades of time. This finding is not what would be predicted (nor is it what we expected at the start of the research project[8]) from any version of the simple hypothesis that rising temperatures act to increase natural disasters.

What is instead striking, in both column 1 and column 3, is a strong statistical role played by a variable capturing the variability of temperature -- as here measured by standard deviation. Its coefficient, in the first and third columns of Table 1, is approximately 0.5 and the tightly defined confidence intervals imply that the null of zero can be rejected at the 1% level. The implied 'SD elasticity of disasters' is thus 0.5.

[8] Initially we presumed we had made a coding error.

A natural check on cross-national results is to examine the patterns in data on a large number of regions across the geography of a single large country. Such a check is likely to be particularly useful in a nation with considerable internal variation in climatic conditions. The United States offers that opportunity.

Table 2, using U.S. data, is identical in spirit to Table 1. In this case, however, there are effectively only 49 long-difference observations (one 'long difference' for each of the 49 contiguous states). A strong predictive statistical role is again found for SD Temperature. In column 1 of Table 2, for example, the coefficient in this log-log equation is close to 3.5, which implies a substantial elasticity of disasters with respect to the standard deviation of temperature. It can be seen, in columns 2 and 3, that in Table 2 the level of temperature actually enters negatively and with a small standard error. This is counter to what conventional wisdom might have predicted.

In order to ensure that the particular data set (GDIS) on natural disasters is not somehow creating false patterns, we also checked the U.S. results by using information from a different source, the Federal Emergency Management Agency (FEMA) government database. Total disasters in that data set are measured slightly differently; the data, for example, include an explicit count of hurricanes and do not include a measure of droughts. Moreover, and valuably for the current paper's analysis, these disasters are judged as major by the federal agency and FEMA data are likely to be less prone to time bias (greater reporting, for instance) that might affect GDIS data.

The regression structure estimates the same form of long-difference equation as in Table 1 and Table 2, but now using FEMA data on United States disasters across 49 states.

The results are in Table 3. They appear encouragingly consistent with those in the earlier tables. Once again, it is possible to reject the null of zero on SD Temperature and that variable enters positively. The implied elasticity in columns 1 and 3 is a little below unity.

Again, in the last two columns of Table 3, there is no clear role for the level of temperature as a predictor of more natural disasters and the point estimate is negative.

Figures 1-9 illustrate these kinds of results in a simple graphical way. They also provide information about the underlying aggregate patterns in the data. For our objectives to be statistically feasible, temperature variability must have a substantial degree of orthogonality from mean temperature itself. The time-series patterns in temperature and temperature variability in our data do have that characteristic. Raw numbers are shown as line graphs in Figures 5 and 6. These depict, for our two data sets, the mean of temperature and the standard deviation of temperature. This is for (i) a set of countries using the years available for World Values Survey data and (ii) for the United States using the period available in U.S. Behavioral Risk Factor Surveillance System data that we use later. The numbers in the Figures are in Celsius and based on 10-year calculations (so, importantly, Figure 5 and Figure 6 give the mean temperature in Celsius over ten years and the standard deviation of temperature in Celsius over ten years). Presenting the information with ten-year smoothing helps to show the underlying patterns and offers a middle ground between the 30-year averaging in the analysis and the fluctuations inevitable in annual data.

Some of the exact patterns in Figure 5 may not be widely known. It can be seen that, despite considerable fluctuations, average global temperature for the WVS countries has been broadly trending up, in a secular sense, over time. That is consistent with what we know about a warming planet. However, and in contrast, 10-year temperature variability does not behave in the same way. As proxied by the standard deviation of temperature, the variability has been neither consistently rising nor falling through the years. For example, the jagged series in Figure 5 reveals a substantial fall in temperature variability -- a kind of rough V shape is visible in the standard deviation -- during much of the first decade of the 2000s.

Figure 6, which is for the United States alone, exhibits a similar V-shaped pattern for the standard deviation of temperature (perhaps thereby indicating that temperature variability has a common world component). The U.S. time trend in mean temperature is a slightly rising one, in the data of Figure 6, but its erratic nature is noticeable.

**Results for Different Kinds of Disasters**

We now disaggregate. Table 4 extends the previous cross-national analysis on total disasters. Its first three columns give regression equations for storms, floods, and droughts. For ease of comparison, the final column in Table 4 is a reminder of an earlier overall equation for total disasters.

Columns 1 and 2 of Table 4 reveal, across 176 countries, that SD Temperature enters positively (coefficient 0.441 and coefficient 0.137, respectively), and with a small standard error, as predictors of storms and floods. Column 3 is different. Here the standard deviation of temperature has a negative, although very small, implied elasticity, and it is possible to reject the null of zero at the 10% significance level. Overall, there continues to be no discernible evidence -- in any of the columns of Table 4 -- that higher temperatures are associated with a greater chance of natural disasters.

Table 5, using data on the states of the United States, is again similar, although not exactly so. Once more, the first three columns report long-difference equations for storms, floods, and droughts. No clear conclusions can be drawn from the third column on droughts. In the other columns of Table 5 the results are approximately consistent with the idea that SD Temperature is associated with more storms and floods (though the positive point estimate in the second column is not well defined) and that Mean Temperature has the wrong sign by the standards of conventional wisdom.

Table 6 switches to the FEMA measures of disasters in the United States, and this allows separate equations for storms, floods, and hurricanes. Perhaps the most notable findings, when compared to Table 5, are that a hurricanes coefficient is large at 0.582 and that higher temperatures here (in column 2 of Table 5) are now associated with a greater risk of floods. The storms coefficient, and therefore implied elasticity, on SD Temperature in the first column of Table 6 is substantial at 0.733, but the null of zero cannot be rejected at conventional significance cut-off levels.

### Human Wellbeing and Disasters

What are the societal implications of the disasters equations above? In an attempt to try to assess one important aspect of that, we turn to measures of psychological well-being.

To begin to test for a disasters-centered effect, Tables 7-8 use international data on the World Values Survey (WVS) sample of countries and periods. Data are available for 93 countries over the period 1990-2018. Table 7 uses life satisfaction scores as its dependent variable (those, as answered by respondents in the survey, are coded from 1 to 10). It is possible to object to that, although a huge literature in social science has drawn on such data, and Kaiser-Oswald 2022 demonstrate the value of cardinality as a powerful predictor of actions following a longstanding Milton Friedman-ite methodological spirit). Table 8 uses happiness scores as its dependent variable (those, as answered by respondents in the survey, are coded on a four-point scale from 0 to 3). Happiness can be viewed more as a measure of affective well-being that captures an individual's usual feelings and emotions (Pavot and Diener, 1993).

Tables 7 and 8 are annual fixed-effects equations. This is to allow a judgment about the impact on people within the year of a particular natural disaster. The personal independent variables -- to correct for potential confounders -- are the ones largely conventional (for example, Di Tella 2001, 2003) in the statistical literature on the study of human-wellbeing equations: they are age, income, education, sex, marital status, health status, religiosity, number of children, employment status, and (perhaps less conventionally) self-reported honesty.

Both Table 7 and Table 8 include, as extra independent variables, the number of disasters. These are entered in the regression equations as (i) the total number of disasters in the fourth column of the tables and (ii) separately in columns 1, 2, and 3 as numbers of an individual type of disaster. Across the 93 countries, there are now 1393 sub-regions. Year effects and sub-region effects are included on the right hand side of the equations; so this is effectively a 28-years times 1393-regions fixed-effects approach.

As would be expected, natural disasters reduce people's feelings of life satisfaction and happiness. Each of the eight estimated coefficients in Table 7 and Table 8 is negative. In Table 7, the null of zero can be rejected at conventional confidence levels in three of the four columns (floods, in column 2, being the exception). The same is true in Table 8.

The disasters variables in Tables 7-8 are entered in logarithmic form. Hence it is appropriate to calculate for both what might be termed the elasticity of wellbeing with respect to disasters. To do that, it is necessary to divide the key coefficient by the mean of the dependent variable (approximately 6.7 in the case of life satisfaction and approximately 2.5 in the case of happiness). It is perhaps most natural to perform that calculation for the final column of each of the two tables. What emerges is a small elasticity of approximately -0.01. In other words, for both kinds of psychological dependent variables, a doubling of the number of disasters would be associated with a drop of one percent in measured psychological wellbeing on a cardinal scale. That result is for the international data. It would imply, from the mean of life satisfaction, a fall of 0.068 points in life satisfaction. To put that into perspective, the ‘I am currently separated’ (in a marital sense) coefficient in the life satisfaction equation is approximately -0.21.

The statistical significance of disasters does not hold true in psychological wellbeing equations for the United States. Tables 9-10 switch to that nation and attempt to check the WVS results using BRFSS data. The data consists of approximately 1.7 million observations from randomly sampled residents of the United States. Here the focus is once again on sub-national fixed effects equations. In these statistical estimates, temperature variables are tried (i) at the county level and (ii) at the state level. Because of the need for data availability on wellbeing measures at the fine geographical-area level, the time period covered in the analysis is now restricted to years 2005-2011. Overall for the USA, in the life satisfaction and happiness equations in Tables 9-10, a variable for disasters enters negatively in only three of eight equations, and is never statistically significantly at standard confidence levels (which is broadly consistent with the ideas of Frijters et al. 2023).

Interestingly, our global and U.S. estimates do not uncover large wellbeing effects -- at the level of wide geographical areas[9] measured over a full year -- from natural disasters.

## Checks

For reasons of conciseness, a large number of extra tables and figures are made available in the online Appendix. To distinguish against those in the main body of the paper, these extra tables and figures are assigned by letter rather than by numerical title.

First, an extended collection of the paper’s main material, both for the countries of the world and for the states of the USA, is given in Appendix Tables A-H and Figures A-P. This includes detailed results for each individual kind of natural disaster (storms, floods, droughts) in the GDIS data set.

[9] It is important to stress that those citizens very intimately affected by the disasters are not the group being measured here.

As is inevitable, the level of statistical power is lower in the case of single-type disasters than for total disasters, and thus the confidence intervals tend to be larger. Nevertheless, one fact about the scatter plots is perhaps worth noting. It is that, in each case of a relationship between disasters and the level of temperature, the relevant scatter plot has a negative gradient. Hence, even the point estimates in simple graphs are not consistent with the idea that regions with faster warming have growing numbers of (individual kinds of) disasters.

Second, in occasional parts of the natural-log analysis it was necessary to add epsilon to convert a negative number into a positive one before logs were taken. For this reason, and as a general robustness check, the equivalent set of tables and figures was created using instead the well-known inverse hyperbolic sine (IHS) transformation. This approach has weaknesses as well as advantages (discussed recently in Norton 2022), but IHS has a long history of use in skewed data or with zero or negative values. These alternative results, which reassuringly mirror the earlier logarithm ones, are in the Appendix in Tables I-P and Figures A1-B8.

Third, as explained earlier, FEMA data on U.S. disasters are a different source of information and in some cases the count is of different disasters compared to the GDIS data set. Therefore another full set of tables and figures is given in the Appendix. These are titled Tables Q-X and Figures C1-C16. The same broad conclusions continue to emerge in the FEMA equations.

Fourth, however, there is now an example where there is a positive relationship between (long differences in) a type of disaster and mean temperature. This is for floods in Appendix Figure C6, which is in logs, and in Figure C14, which uses the IHS transformation.

Fifth, and as explained earlier, it is not conceptually ideal to study the paper's relationships using standard un-logged data. For completeness, however, those simple results are reported in the Appendix (in Appendix Tables A1-A7) and continue to point to the same broad conclusions.

Sixth, could it be that the main effect is coming through from the top of the temperature distribution alone -- from simply some very high 'extreme' temperatures -- and that it is misleading to view the key predictor as the standard deviation of temperature? We checked this and the extreme highs do not explain the movements in numbers of disasters. An illustration is Figures D1-D4 in the Appendix.

## Conclusions

It is common in public discussion to hear the idea that rising world temperatures -- popularly known as global warming -- are creating a greater number of natural disasters. Our results point to a different conclusion.

The paper presents new evidence consistent with the idea that there is a fundamental influence of temperature variability upon natural disasters (and thereby indirectly, as we demonstrate formally, upon the psychological wellbeing of humans). Most previous research on climate change -- in both formal climate-science and the social sciences -- has stressed instead the role of the level of temperature. We have adjusted, in this paper's regression equations, for mean temperature as well as its variability, but it is the latter variable that comes through as more consistently important.

While the paper's results as described above in the main text are based on logarithmic specifications, we show in the online Appendix that similar findings emerge from inverse hyperbolic sine (IHS) transformations or simple linear[10] unlogged forms. Nor, as mentioned, is the paper's standard-deviation result driven merely by extremes of high temperatures. [11]

The key independent variable used in the long-difference equations is the standard deviation of temperature averaged over 30 years[12]. As will be known by every scholar who works on the topic of carbon emissions and global warming, the research literature in this important area is large. That means it can be difficult to be certain what has previously been shown in one set of research journals or another. To our knowledge, however, this paper's findings are new ones.[13]

The paper also studies the implied psychological-wellbeing losses from natural disasters. In the analysis, when the level of a whole region is examined, the size of these losses is smaller than might be expected. One potential interpretation and consequence of this is that, no matter how painful disasters are for those closely affected, it is conceivable that there might be relatively little political support among voters for expensive mitigation strategies.

What, in a deep sense, explains the variability-disasters patterns we find? The straightforward answer is that we do not know. We hope that other researchers will discover ways to build upon this paper's tables and graphs. One potential avenue might be in some way to pursue the spirit of Jensen's inequality in standard mathematics (the secant line of a convex function[14] lies above the graph of the function). Another possible way forward is that, in the wellbeing-results section, where fairly small negative or even zero coefficients are found, progress might be made by pursuing the interesting distinction in Hsiang (2016) between direct effects and belief effects. A further possibility is that the intriguing U-shape in mortality that has recently been uncovered by Carleton et al. (2022) may offer a potential signpost to an avenue of inquiry.

---

[10] Intuitively, a linear specification is not entirely natural, because 4 more disasters would be a small rise in a place that normally has a 100 disasters a decade but a large rise in a place with 10 disasters a decade.

[11] If low temperatures change very little and high temperatures go up then mechanically the standard deviation of temperature will increase.

[12] As explained in the paper, in most of the analysis the standard deviation SD is across the first block of thirty years, and then we take the difference compared to the SD across the next block of thirty years. So these are 'long differences' in SD. The dependent variable is the long differences in number of disasters.

[13] Judging by our bibliometric searches, across a variety of journals and disciplines, using the Web of Science. We also thank the climate-change modeler David Stainforth for his assessment. Some previous uses of the standard deviation of temperature, in different settings than ours, are mentioned earlier in the paper.

[14] If, for example, outcomes are a convex and increasing function of temperature.

In conclusion, on the balance of the paper's estimates, it is likely that the sheer variability of temperature[15] itself has severe consequences. We believe these issues demand further attention.

[15] This can still be viewed as an example, and side product, of climate change.

## Disaster Equations for 176 Countries and for the U.S. States

**Table 1- Long-Difference Disaster Equations with Temperature Variables 1960 to 2018 [176 countries]**

| | (1)<br>Total Disasters (Log) | (2)<br>Total Disasters (Log) | (3)<br>Total Disasters (Log) |
|---|---|---|---|
| SD Temperature | $0.535^{***}$ | | $0.533^{***}$ |
| (Log) | [0.326,0.744] | | [0.325,0.740] |
| Mean | | -0.066 | -0.065 |
| Temperature | | [-0.158,0.025] | [-0.159,0.028] |
| (Log) | | | |
| No. sub-regions | 2355 | 2355 | 2355 |
| R-squared | 0.01 | 0.005 | 0.02 |
| No. of countries | 176 | 176 | 176 |
| Mean Dep. Var. | 0.02 | 0.02 | 0.02 |
| Std. Dev. | 0.04 | 0.04 | 0.04 |

Taking long differences in natural logarithms, the dependent variable here -- and in later tables except where specified -- is a measure of total disasters. It is the long difference in the average occurrence of total disasters (i.e. the sum of storms, floods, and droughts) between the periods 1960-1989 and 1990-2018 in each sub-national region (in natural logs). Therefore there is in effect a single observation for each sub-national region within each nation.

SD (Mean) Temperature is calculated equivalently as the long difference in the standard deviation (mean) of the temperature reached, between the periods 1960-1989 and 1990-2018, in each sub-national region (in natural logs). Put explicitly, we take the standard deviation SD across the first block of thirty years, and take the difference compared to the SD across the next block of thirty years. So these are 'long differences' in SD.

Measured as long differences in logarithms, the mean and standard deviation of the variable SD Temperature are 0.001 and 0.008, respectively, and of the variable Mean Temperature are 0.025 and 0.041, respectively. Standard errors are clustered at a sub-national region level, and 95% confidence intervals are reported in brackets. The stars *** , **, and * denote significance level at 1%, 5%, & 10%, respectively. The equations' constant terms are not reported.

Source of data on numbers of disasters
Disasters data are taken from the GDIS database: Rosvold, E., and H. Buhaug, (2021). Geocoded Disasters (GDIS) Dataset. *NASA Socioeconomic Data and Applications Center (SEDAC).*

Source of data on temperature and temperature units
All temperature variables are taken from the CRU database*, as explained in the paper, and are measured in Celsius units.

*Harris, I.C., P.D. Jones, and T. Osborn, (2020). CRU TS4.04: Climatic Research Unit (CRU) Time Series (TS) version 4.04 of high resolution gridded data of month by month variation in climate (Jan. 1901 Dec. 2019). *Centre for Environmental Data Analysis*

**Table 2- Long-Difference Disaster Equations with Temperature Variables United States 1960 to 2018 [49 U.S. states]**

| | (1)<br>Total Disasters (Log) | (2)<br>Total Disasters (Log) | (3)<br>Total Disasters (Log) |
|---|---|---|---|
| SD Temperature (Log) | 3.427*** | | 3.700*** |
| | [0.965,5.889] | | [1.693,5.707] |
| Mean Temperature (Log) | | -7.728*** | -7.954*** |
| | | [-10.694,-4.763] | [-10.935,-4.973] |
| No. states | 49 | 49 | 49 |
| R-squared | 0.10 | 0.28 | 0.40 |
| Mean Dep. Var. | 0.48 | 0.48 | 0.48 |
| Std. Dev. | 0.27 | 0.27 | 0.27 |

Taking long differences in logarithms, the dependent variable is a measure of total disasters. It is the long difference in the average occurrence of total disasters (i.e. the sum of storms, floods, and droughts) between the periods 1960-1989 and 1990-2018 in each of the contiguous states of the USA (in natural logs). There is therefore a single observation for each state within the nation. SD (Mean) Temperature is calculated equivalently as the long difference in the standard deviation (mean) of the temperature reached, between the periods 1960-1989 and 1990-2018, in each state (in natural logs). The sample excludes Alaska, Hawaii and US overseas territories. Measured as long differences in logarithms, the mean and standard deviation of the variable SD Temperature are 0.051 and 0.024, respectively, and of the variable Mean Temperature are 0.054 and 0.018, respectively. Standard errors are clustered at the state level, and 95% confidence intervals are reported in brackets. *** , **, and * denote significance level at 1%, 5%, & 10%, respectively.

Source for the USA disasters data is also Rosvold and Buhaug (2021).

All USA temperature variables are taken from the CRU database, as explained in the paper, and are measured in Celsius units.

**Table 3 - Total Disaster Equations with Temperature Variables United States 1960 to 2018 using FEMA data [49 U.S. states]**

| | (1)<br>Total Disasters (Log) | (2)<br>Total Disasters (Log) | (3)<br>Total Disasters (Log) |
|---|---|---|---|
| SD Temperature | 0.861** | | 0.857** |
| (Log) | [0.110,1.612] | | [0.091,1.622] |
| Mean Temperature | | -0.331 | -0.254 |
| (Log) | | [-2.707,2.045] | [-2.547,2.039] |
| No. states | 49 | 49 | 49 |
| R-squared | 0.07 | 0.002 | 0.07 |
| Mean Dep. Var. | 0.36 | 0.36 | 0.36 |
| Std. Dev. | 0.18 | 0.18 | 0.18 |

The dependent variable is total disasters which measure the long difference in the average occurrence of the total disasters (i.e. the sum of storms, floods, and hurricanes) between the periods 1960-1989 and 1990-2018 in each state (in natural logs). SD (Mean) Temperature is calculated as the long difference in the standard deviation (mean) of the temperature reached, between the periods 1960-1989 and 1990-2018 in each state (in natural logs). The sample excludes Alaska, Hawaii and US overseas territories. The mean and standard deviation of the variable SD Temperature are 0.115 and 0.055, respectively, and of the variable Mean Temperature are 0.065 and 0.025, respectively. Standard errors are clustered at the state level, and 95% confidence intervals are reported in brackets. *** , **, and * denote significance level at 1%, 5%, & 10%, respectively.

**Table 4- Long-Difference Disaster Equations with Temperature Variables 1960 to 2018 [176 countries]**

| | (1) Storms (Log) | (2) Floods (Log) | (3) Droughts (Log) | (4) Total Disasters (Log) |
|---|---|---|---|---|
| SD Temperature (Log) | 0.441*** | 0.137*** | -0.022* | 0.533*** |
| | [0.289,0.593] | [0.039,0.235] | [-0.044,0.000] | [0.325,0.740] |
| Mean Temperature (Log) | -0.012 | -0.042 | -0.010 | -0.065 |
| | [-0.032,0.007] | [-0.104,0.020] | [-0.023,0.003] | [-0.159,0.028] |
| No. sub-regions | 2355 | 2355 | 2355 | 2355 |
| R-squared | 0.02 | 0.01 | 0.01 | 0.02 |
| No. of countries | 176 | 176 | 176 | 176 |
| Mean Dep. Var. | 0.01 | 0.01 | 0.006 | 0.02 |
| Std. Dev. | 0.02 | 0.02 | 0.01 | 0.04 |

Here the dependent variables on storms, floods, droughts, and total disasters measure the long difference in the average occurrence of each disaster between the periods 1960-1989 and 1990-2018 in each sub-national region (in natural logs). SD (Mean) Temperature is calculated equivalently as the long difference in the standard deviation (mean) of the temperature reached, between the periods 1960-1989 and 1990-2018 in each sub-national region (in natural logs). The mean and standard deviation of the variable SD Temperature are 0.001 and 0.008, respectively, and of the variable Mean Temperature are 0.025 and 0.041, respectively. Standard errors are clustered at the sub-national region level, and 95% confidence intervals are reported in brackets. *** , **, and * denote significance level at 1%, 5%, & 10%, respectively.

**Table 5- Long-Difference Disaster Equations with Temperature Variables United States 1960 to 2018 [49 U.S. states]**

| | (1) Storms (Log) | (2) Floods (Log) | (3) Droughts (Log) | (4) Total Disasters (Log) |
|---|---|---|---|---|
| SD Temperature (Log) | 3.883*** | 0.283 | -0.055 | 3.700*** |
| | [2.081,5.684] | [-0.467,1.032] | [-0.200,0.091] | [1.693,5.707] |
| Mean Temperature (Log) | -7.193*** | -2.191*** | -0.165 | -7.954*** |
| | [-9.947,-4.438] | [-3.473,-0.909] | [-0.405,0.075] | [-10.935,-4.973] |
| No. states | 49 | 49 | 49 | 49 |
| R-squared | 0.41 | 0.19 | 0.05 | 0.40 |
| Mean Dep. Var. | 0.41 | 0.12 | 0.02 | 0.48 |
| Std. Dev. | 0.25 | 0.09 | 0.02 | 0.27 |

Taking long differences in logarithms, the dependent variables on storms, floods, droughts, and total disasters measure the long difference in the average occurrence of each disaster between the periods 1960-1989 and 1990-2018 in each state (in natural logs). SD (Mean) Temperature is calculated equivalently as the long difference in the standard deviation (mean) of the temperature reached between the periods 1960-1989 and 1990-2018 in each state (in natural logs). The sample excludes Alaska, Hawaii, and US overseas territories. Measured as long differences in logarithms, the mean and standard deviation of the variable SD Temperature are 0.051 and 0.024, respectively, and of the variable Mean Temperature are 0.054 and 0.018, respectively. Standard errors are clustered at state level, and 95% confidence intervals are reported in brackets. *** , **, and * denote significance level at 1%, 5%, & 10%, respectively.

**Table 6 - Disaster Equations with Temperature Variables United States 1960 to 2018 using FEMA data [49 U.S. states]**

| | (1)<br>Storms (Log) | (2)<br>Floods (Log) | (3)<br>Hurricanes (Log) | (4)<br>Total Disasters (Log) |
|---|---|---|---|---|
| SD Temperature | 0.733 | -0.263 | 0.582** | 0.857** |
| (Log) | [-0.189,1.656] | [-0.702,0.176] | [0.099,1.065] | [0.091,1.622] |
| Mean Temperature | -0.568 | 2.058*** | -2.041*** | -0.254 |
| (Log) | [-2.687,1.551] | [0.841,3.275] | [-3.529,-0.554] | [-2.547,2.039] |
| No. states | 49 | 49 | 49 | 49 |
| R-squared | 0.05 | 0.20 | 0.22 | 0.07 |
| Mean Dep. Var. | 0.44 | -0.11 | 0.12 | 0.36 |
| Std. Dev. | 0.20 | 0.12 | 0.13 | 0.18 |

The dependent variables on storms, floods, hurricanes, and total disasters measure the long difference in the average occurrence of each disaster between the periods 1960-1989 and 1990-2018 in each state (in natural logs). SD (Mean) Temperature is calculated as the long difference in the standard deviation (mean) of the temperature reached between the periods 1960-1989 and 1990-2018 in each state (in natural logs). The sample excludes Alaska, Hawaii, and US overseas territories. The mean and standard deviation of the variable SD Temperature are 0.115 and 0.055, respectively, and of the variable Mean Temperature are 0.065 and 0.025, respectively. Standard errors are clustered at state level, and 95% confidence intervals are reported in brackets. *** , **, and * denote significance level at 1%, 5%, & 10%, respectively.

## Happiness and Life Satisfaction Equations with Disaster Variables: Fixed Effects Equations

**Table 7 - Life Satisfaction Equations with Disaster Variables – World Values Survey Data 1990 to 2018 [93 countries]**

| | (1) | (2) | (3) | (4) |
|---|---|---|---|---|
| Storm (Log) | -0.133** | | | |
| | [-0.260,-0.005] | | | |
| Flood (Log) | | -0.025 | | |
| | | [-0.119,0.069] | | |
| Drought (Log) | | | -0.288** | |
| | | | [-0.546,-0.030] | |
| Total disasters (Log) (i.e. their sum) | | | | -0.076* |
| | | | | [-0.155,0.002] |
| Observations | 228111 | 228111 | 228111 | 228111 |
| R-squared | 0.28 | 0.28 | 0.28 | 0.28 |
| No. of sub-national regions | 1393 | 1393 | 1393 | 1393 |
| Mean Dep. Var. | 6.68 | 6.68 | 6.68 | 6.68 |
| Std. Dev. | 2.39 | 2.39 | 2.39 | 2.39 |
| No. of countries | 93 | 93 | 93 | 93 |

Sub-national region fixed effects and year dummies are included in all columns. The dependent variable is the WVS question "Rate your life satisfaction" (Scale: 1-10). Storms, floods, droughts, and total disasters measure the annual number of occurrences of each disaster in each sub-national region (in natural logs). Standard errors are clustered at the sub-national region level, are reported in parentheses. ***, **, and * denote significance level at 1%, 5%, & 10%, respectively. The baseline controls include age dummies, income, education, sex, marital status, health status, religiosity, number of children, employment status and self-reported honesty.

**Table 8 - Happiness Equations with Disaster Variables – World Values Survey Data 1990 to 2018 [93 countries]**

| | (1) | (2) | (3) | (4) |
|---|---|---|---|---|
| Storm (Log) | -0.039$^{*}$ | | | |
| | [-0.082,0.005] | | | |
| Flood (Log) | | -0.015 | | |
| | | [-0.039,0.010] | | |
| Drought (Log) | | | -0.176$^{***}$ | |
| | | | [-0.271,-0.080] | |
| Total disasters (Log) | | | | -0.032$^{***}$ |
| (i.e. their sum) | | | | [-0.056,-0.009] |
| Observations | 228570 | 228570 | 228570 | 228570 |
| R-squared | 0.26 | 0.26 | 0.26 | 0.26 |
| No. of sub-national regions | 1393 | 1393 | 1393 | 1393 |
| Mean Dep. Var. | 2.09 | 2.09 | 2.09 | 2.09 |
| Std. Dev. | 0.74 | 0.74 | 0.74 | 0.74 |
| No. of countries | 93 | 93 | 93 | 93 |

Sub-national region fixed effects and year dummies are included in all columns. The dependent variable is the WVS question "How happy are you" (Scale: 0-3). Storms, floods, droughts, and total disasters measure the annual number of occurrences of each disaster in each sub-national region (in natural logs). Standard errors are clustered at the sub-national region level, are reported in parentheses. ***, **, and * denote significance level at 1%, 5%, & 10%, respectively. The baseline controls include age dummies, income, education, sex, marital status, health status, religiosity, number of children, employment status and self-reported honesty.

**Table 9 - Life Satisfaction Equations with Disaster Variables – United States Data 2005 to 2011 [2357 U.S. counties]**

| | (1) | (2) | (3) | (4) |
|---|---|---|---|---|
| Storm (Log) | 0.004 | | | |
| | [-0.002,0.010] | | | |
| Flood (Log) | | -0.006 | | |
| | | [-0.016,0.004] | | |
| Drought (Log) | | | -0.013 | |
| | | | [-0.032,0.007] | |
| Total disasters (Log) | | | | 0.000 |
| | | | | [-0.005,0.005] |
| Observations | 1744100 | 1744100 | 1744100 | 1744100 |
| R-squared | 0.18 | 0.18 | 0.18 | 0.18 |
| No. of counties | 2357 | 2357 | 2357 | 2357 |
| Mean Dep. Var. | 3.39 | 3.39 | 3.39 | 3.39 |
| Std. Dev. | 0.63 | 0.63 | 0.63 | 0.63 |
| No. of states | 49 | 49 | 49 | 49 |

County fixed effects and year dummies are included in all columns. The dependent variable is the BRFSS question "In general, how satisfied are you with your life?" (Scale: 1-4). Storms, floods, droughts, and total disasters measure the annual number of occurrences of each disaster in each county (in natural logs). The sample excludes Alaska, Hawaii, and US overseas territories. Standard errors are clustered at the county level, and 95% confidence intervals are reported in parentheses. ***, **, and * denote significance level at 1%, 5%, & 10%, respectively. The baseline controls include age dummies, income, education, sex, marital status, health status, number of children, and employment status.

**Table 10 - Life Satisfaction Equations with Disaster Variables – United States Data 2005 to 2011 [49 U.S. states]**

| | (1) | (2) | (3) | (4) |
|---|---|---|---|---|
| Storm (Log) | 0.002 | | | |
| | [-0.003,0.007] | | | |
| Flood (Log) | | -0.005 | | |
| | | [-0.015,0.005] | | |
| Drought (Log) | | | 0.009 | |
| | | | [-0.010,0.028] | |
| Total disasters (Log) | | | | 0.000 |
| | | | | [-0.005,0.005] |
| Observations | 1744104 | 1744104 | 1744104 | 1744104 |
| R-squared | 0.18 | 0.18 | 0.18 | 0.18 |
| No. of states | 49 | 49 | 49 | 49 |
| Mean Dep. Var. | 3.39 | 3.39 | 3.39 | 3.39 |
| Std. Dev. | 0.63 | 0.63 | 0.63 | 0.63 |

State fixed effects and year dummies are included in all columns. The dependent variable is the BRFSS question "In general, how satisfied are you with your life?" (Scale: 1-4). Storms, floods, droughts, and total disasters measure the annual number of occurrences of each disaster in each state (in natural logs). The sample excludes Alaska, Hawaii, and US overseas territories. Standard errors are clustered at the state level, and 95% confidence intervals are reported in parentheses. ***, **, and * denote significance level at 1%, 5%, & 10%, respectively. The baseline controls include age dummies, income, education, sex, marital status, health status, number of children, and employment status.

# Figures

<u>Which temperature variables predict natural disasters?</u>

## USA (by state)

**Total Disasters**

**Figure 1 – USA Temperature Standard Deviation and Total Disasters Using GDIS Data**

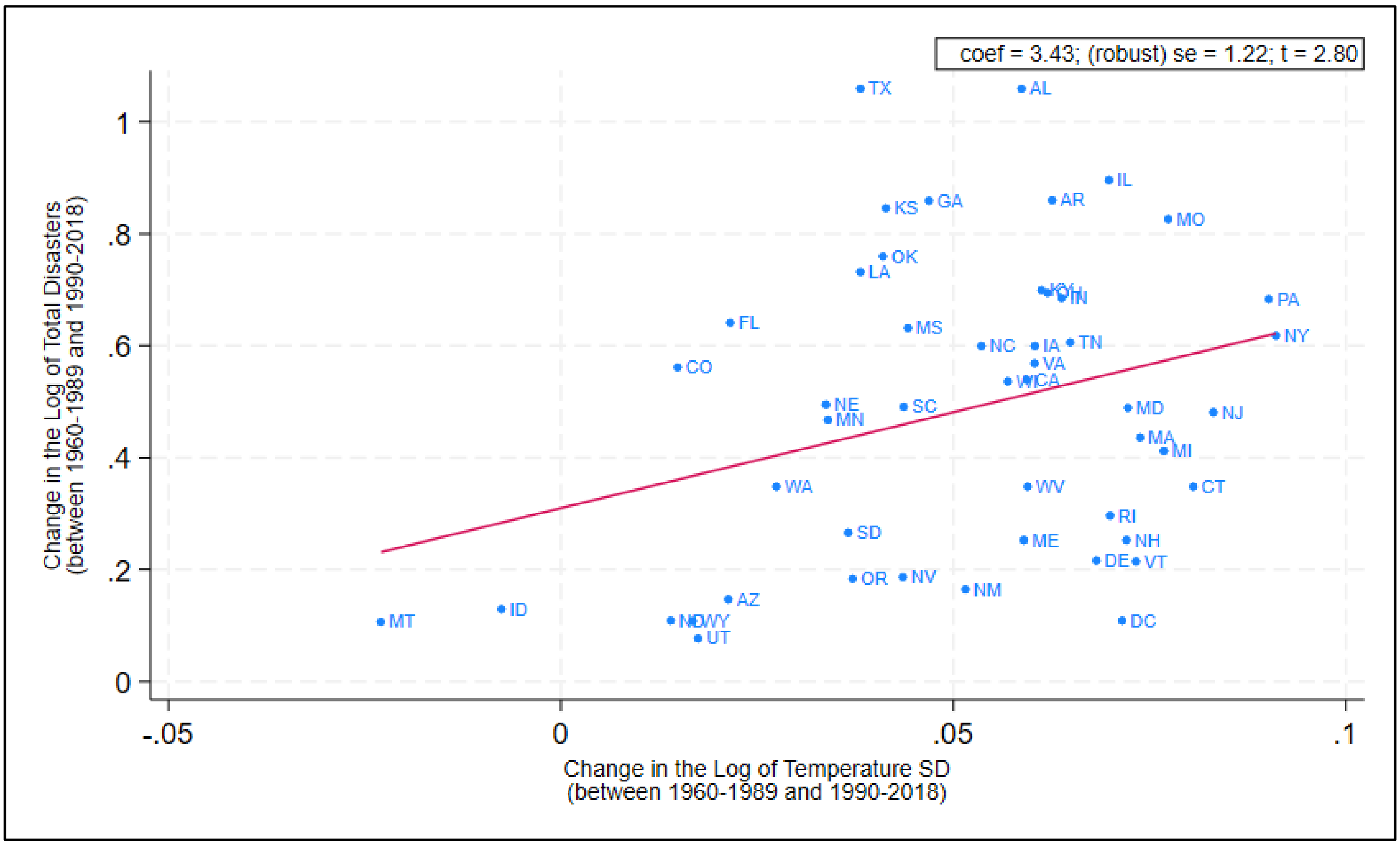

Total disasters measure the long difference in the average occurrence of all disasters between the periods 1960-1989 and 1990-2018 in each state (in natural logs). SD Temperature is calculated as the long difference in the standard deviation of the temperature reached, between the periods 1960-1989 and 1990-2018 in each US state (in natural logs). The sample excludes Alaska, Hawaii, and US overseas territories.

<u>Source of data on numbers of disasters</u>
Disasters data are taken from the GDIS database: Rosvold, E., and H. Buhaug, (2021). Geocoded Disasters (GDIS) Dataset. *NASA Socioeconomic Data and Applications Center (SEDAC).*

<u>Source of data on temperature</u>
All temperature variables here are taken from the CRU database*, as explained in the paper, and are measured in Celsius units.

*Harris, I.C., P.D. Jones, and T. Osborn, (2020). CRU TS4.04: Climatic Research Unit (CRU) Time Series (TS) version 4.04 of high resolution gridded data of month by month variation in climate (Jan. 1901 Dec. 2019). *Centre for Environmental Data Analysis*

**Figure 2 - USA Temperature Mean and Total Disasters Using GDIS Data**

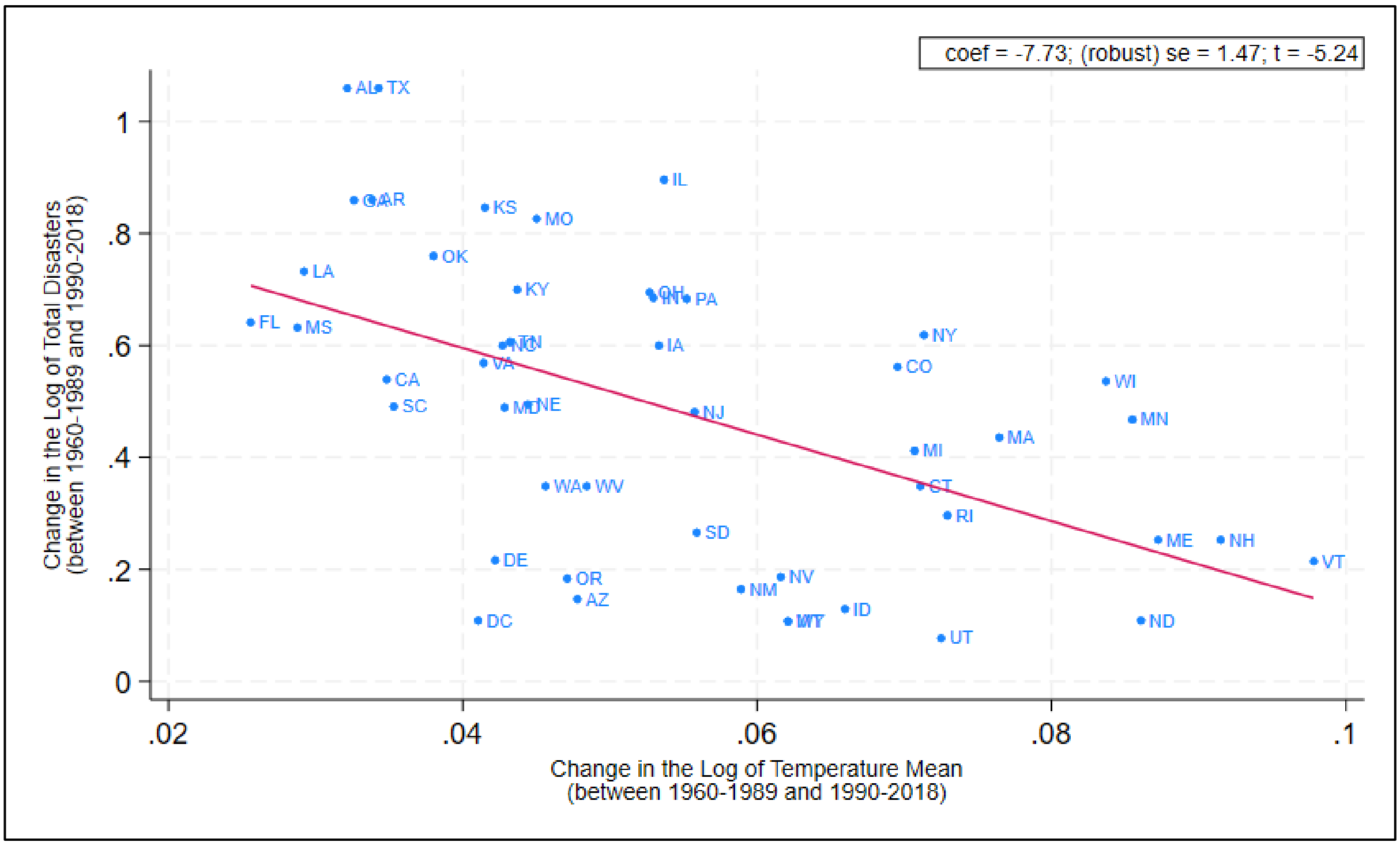

Total disasters measure the long difference in the average occurrence of all disasters between the periods 1960-1989 and 1990-2018 in each state (in natural logs). Mean Temperature is calculated as the long difference in the mean of the temperature reached, between the periods 1960-1989 and 1990-2018 in each state (in natural logs). The sample excludes Alaska, Hawaii, and US overseas territories.

**Figure 3 – USA Temperature Standard Deviation and Total Disasters Using FEMA data**

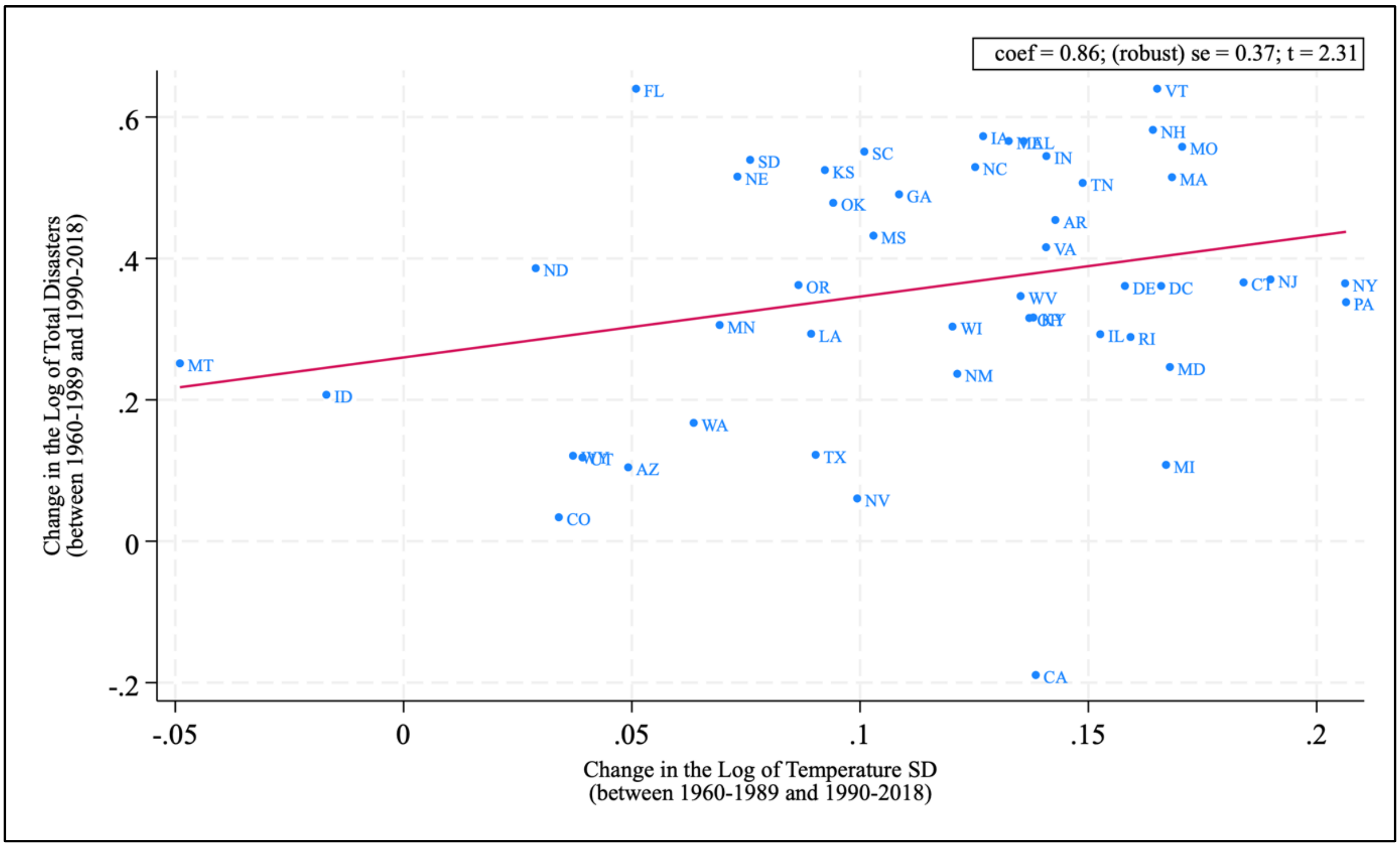

Total disasters measure the long difference in the average occurrence of all disasters between the periods 1960-1989 and 1990-2018 in each state (in natural logs). SD Temperature is calculated as the long difference in the standard deviation of the temperature reached, between the periods 1960-1989 and 1990-2018 in each state (in natural logs). The sample excludes Alaska, Hawaii, and US overseas territories.

<u>Source of data on numbers of disasters</u>
Source for the USA disasters data is the FEMA database*.

Federal Emergency Management Agency (2021). OpenFEMA Dataset, *Retrieved from:* https://www.fema.gov/disaster

<u>Source of data on temperature</u>
All temperature variables are taken from the CRU database*, as explained in the paper, and are measured in Celsius units.

*Harris, I.C., P.D. Jones, and T. Osborn, (2020). CRU TS4.04: Climatic Research Unit (CRU) Time Series (TS) version 4.04 of high resolution gridded data of month by month variation in climate (Jan. 1901 Dec. 2019). *Centre for Environmental Data Analysis*

**Figure 4 - USA Temperature Mean and Total Disasters Using FEMA data**

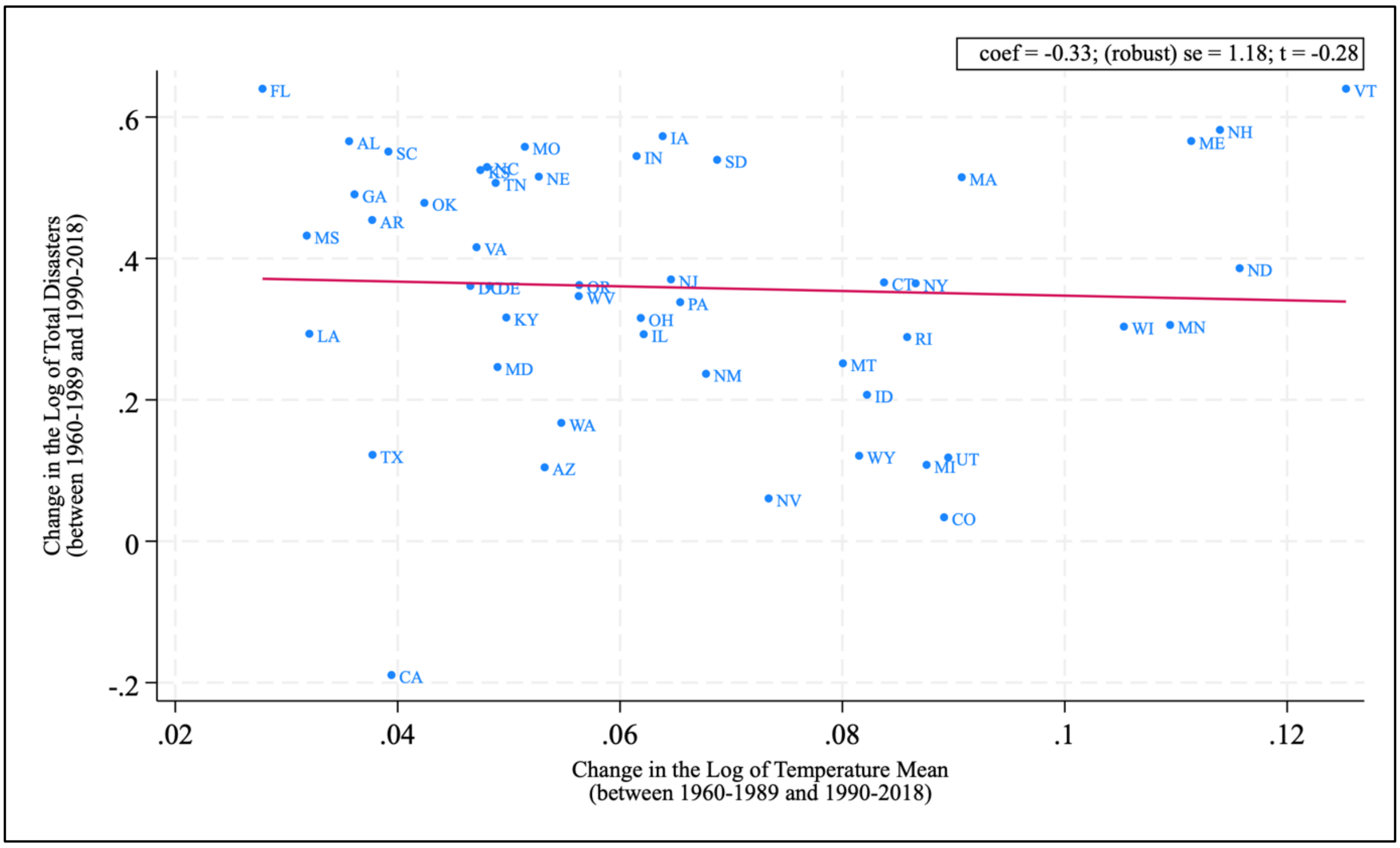

Total disasters measure the long difference in the average occurrence of all disasters between the periods 1960-1989 and 1990-2018 in each state (in natural logs). Mean Temperature is calculated as the long difference in the mean of the temperature reached, between the periods 1960-1989 and 1990-2018 in each state (in natural logs). The sample excludes Alaska, Hawaii, and US overseas territories.

# World (by country)

**Total Disasters**

**Figure 5 – Global Sample: Temperature Standard Deviation and Total Disasters**

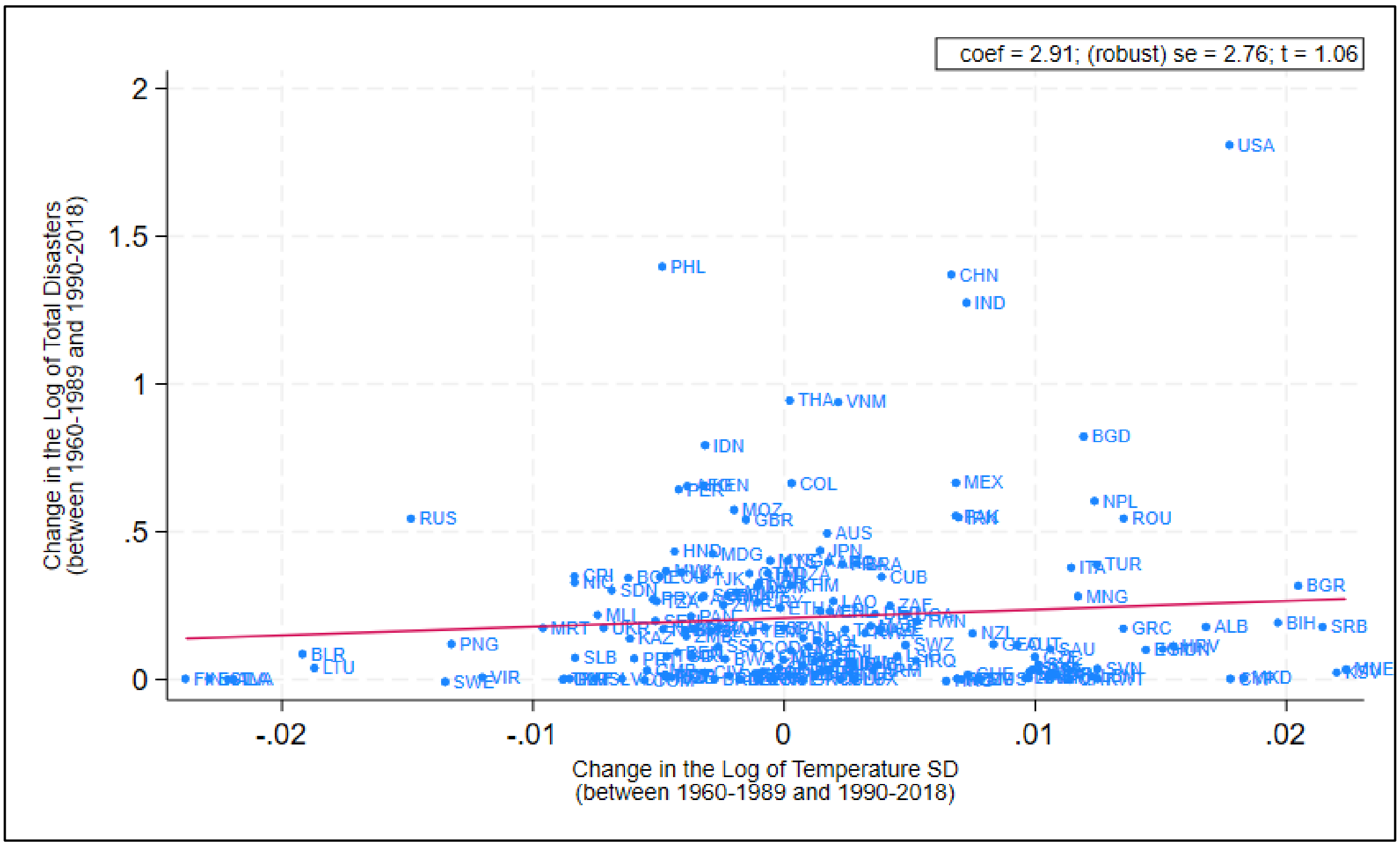

Total disasters measure the long difference in the average occurrence of all disasters between the periods 1960-1989 and 1990-2018 in each country (in natural logs). SD Temperature is calculated as the long difference in the standard deviation of the temperature reached, between the periods 1960-1989 and 1990-2018 in each country (in natural logs).

<u>Source of data on numbers of disasters</u>
Disasters data are taken from the GDIS database: Rosvold, E., and H. Buhaug, (2021). Geocoded Disasters (GDIS) Dataset. *NASA Socioeconomic Data and Applications Center (SEDAC).*

<u>Source of data on temperature</u>
All temperature variables are taken from the CRU database*, as explained in the paper, and are measured in Celsius units.

*Harris, I.C., P.D. Jones, and T. Osborn, (2020). CRU TS4.04: Climatic Research Unit (CRU) Time Series (TS) version 4.04 of high resolution gridded data of month by month variation in climate (Jan. 1901 Dec. 2019). *Centre for Environmental Data Analysis*

**Figure 6 - Global Sample: Temperature Mean and Total Disasters**

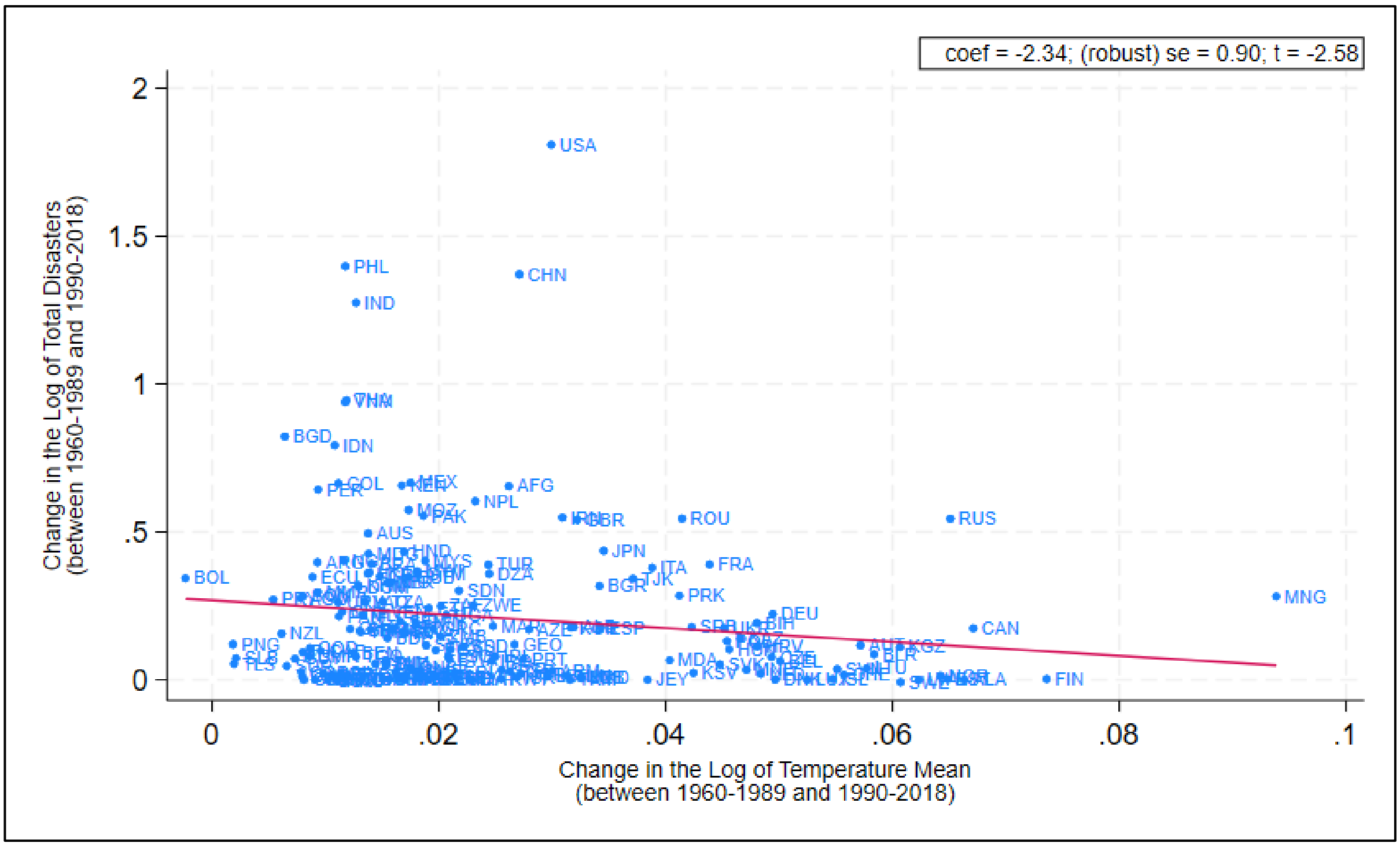

Total disasters measure the long difference in the average occurrence of all disasters between the periods 1960-1989 and 1990-2018 in each country (in natural logs). Mean Temperature is calculated as the long difference in the mean of the temperature reached, between the periods 1960-1989 and 1990-2018 in each country (in natural logs).

<u>Source of data on numbers of disasters</u>

Disasters data are taken from the GDIS database: Rosvold, E., and H. Buhaug, (2021). Geocoded Disasters (GDIS) Dataset. *NASA Socioeconomic Data and Applications Center (SEDAC).*

<u>Source of data on temperature</u>

All temperature variables are taken from the CRU database*, as explained in the paper, and are measured in Celsius units.

*Harris, I.C., P.D. Jones, and T. Osborn, (2020). CRU TS4.04: Climatic Research Unit (CRU) Time Series (TS) version 4.04 of high resolution gridded data of month by month variation in climate (Jan. 1901 Dec. 2019). *Centre for Environmental Data Analysis*

**Figure 5: World Temperature Mean and Temperature Standard Deviation (both calculated over 10 years) for the Nations Covered in the World Values Survey (WVS) Sample.**

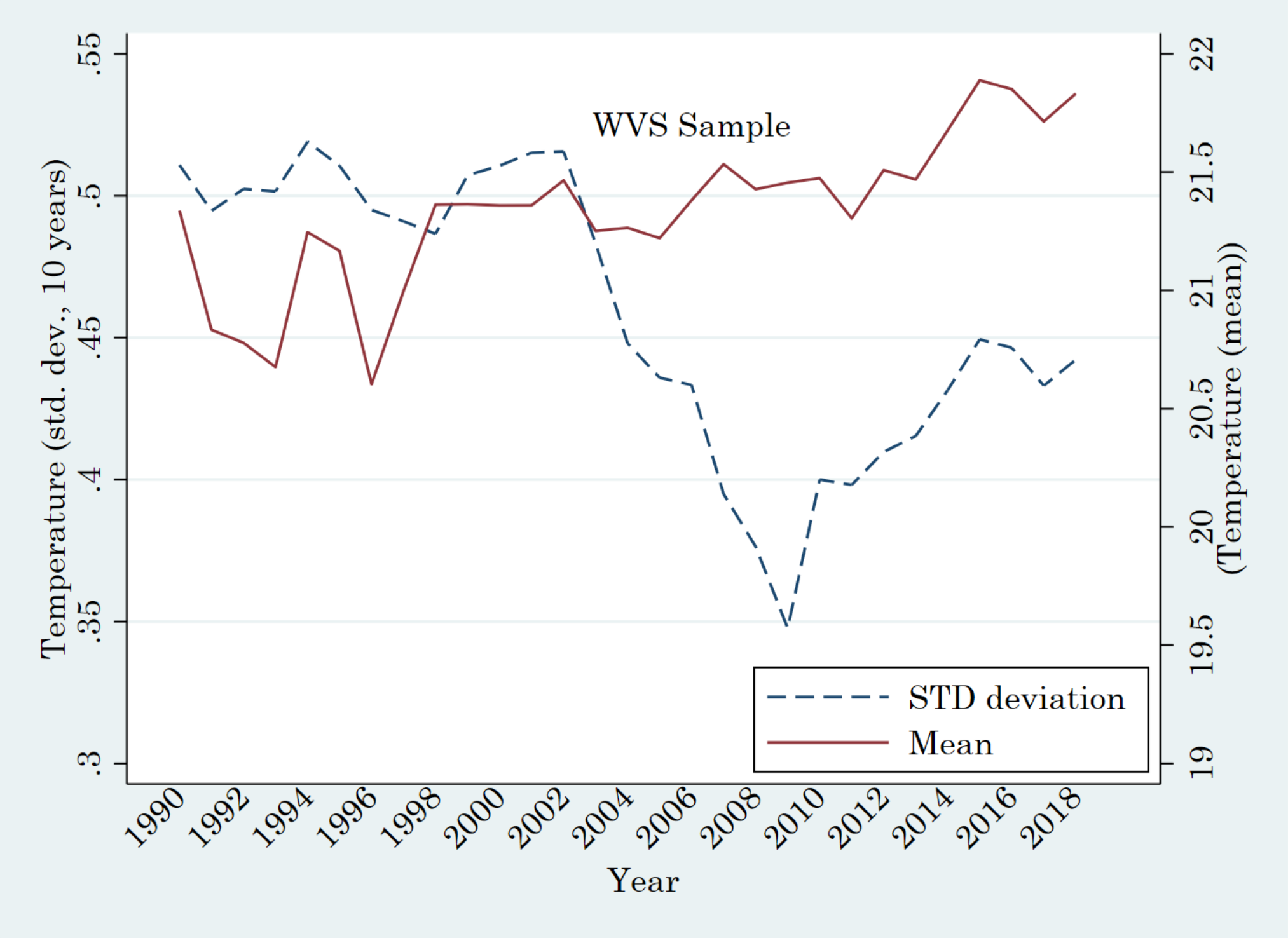

Data source for both figures is the CRU database, as explained in the paper. Temperatures are in Celsius units.

**Figure 6: USA Temperature Mean and Temperature Standard Deviation (both calculated over 10 years)**

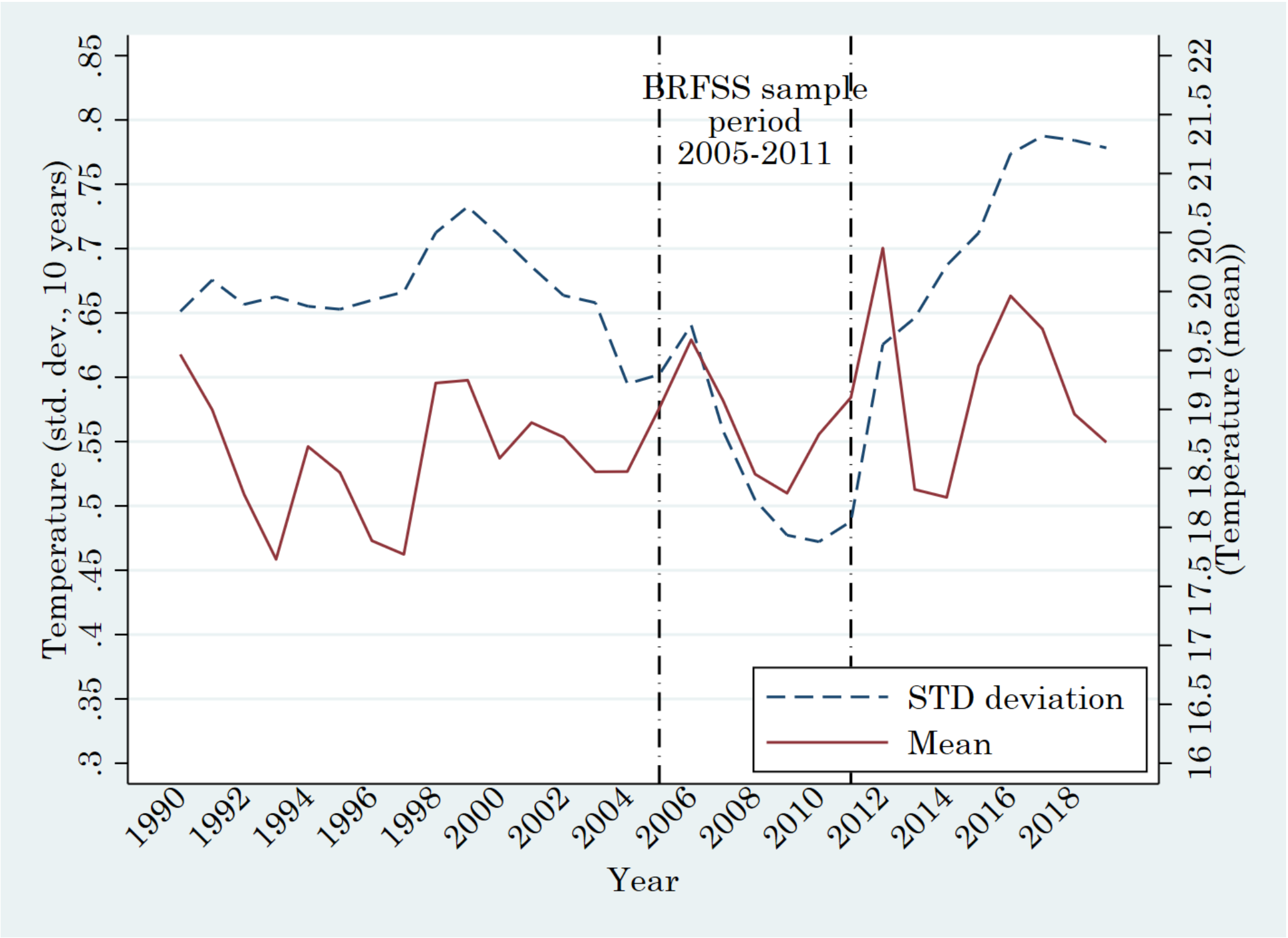

Data source for both figures is the CRU database, as explained in the paper. Temperatures are in Celsius units.

**Figure 7:** Global Total Disasters

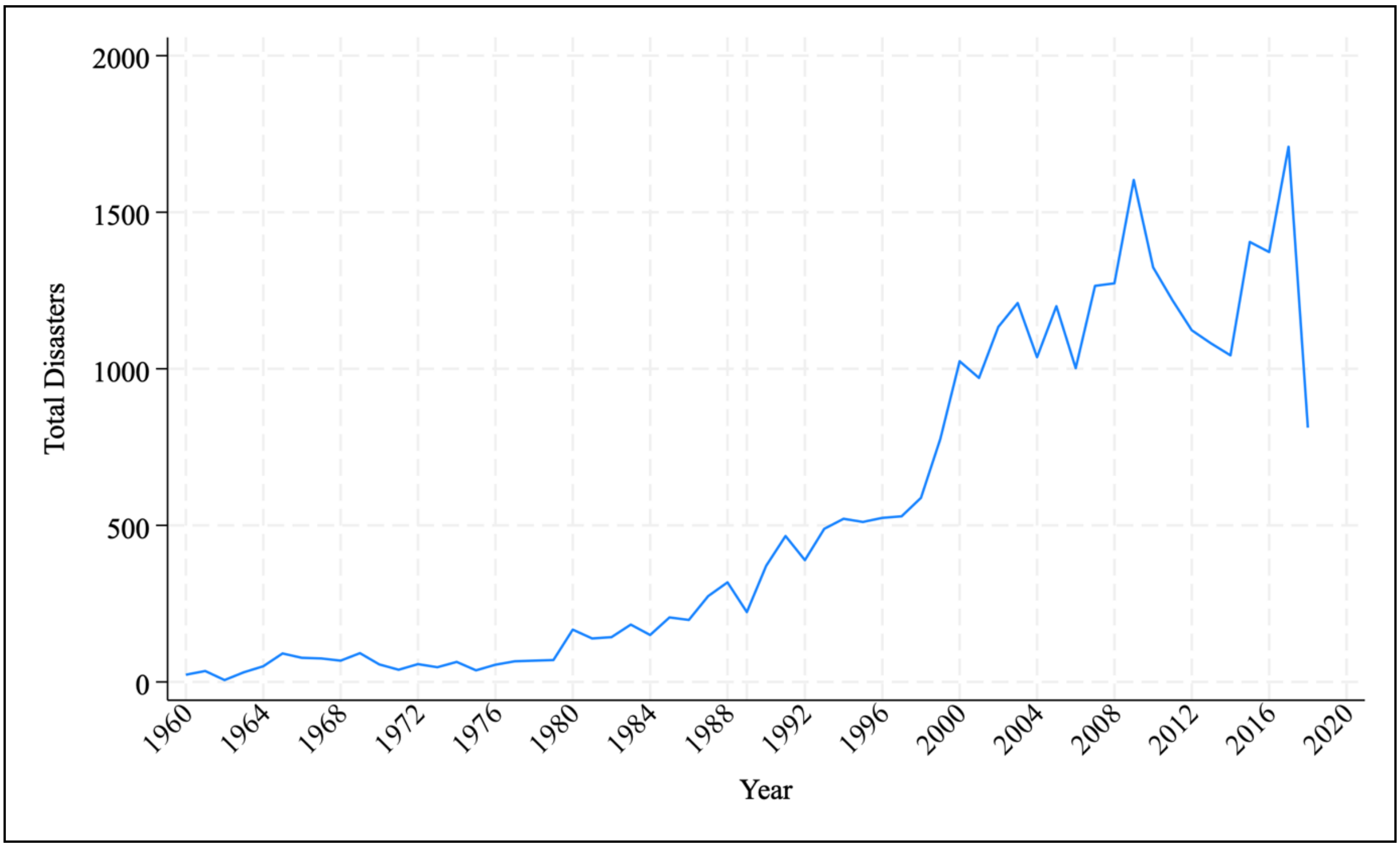

Total disasters measure the annual total number of disasters in the 176 countries of our global sample.

**Figure 8:** Total Disasters in US states – GDIS dataset

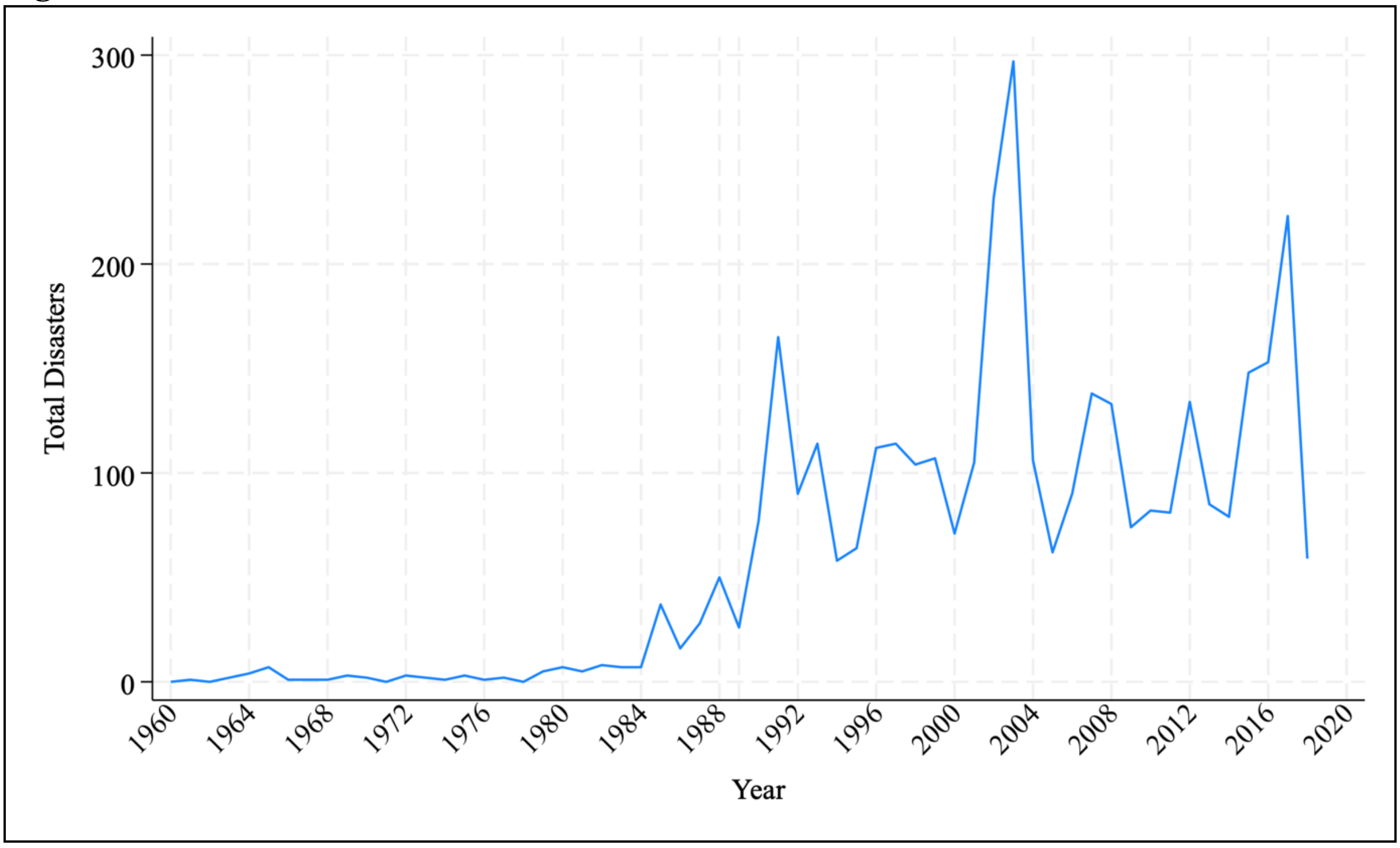

Total disasters measure the annual total number of disasters in the 49 states of our US sample using the GDIS dataset.

**Figure 9:** Total Disasters in US states – FEMA dataset

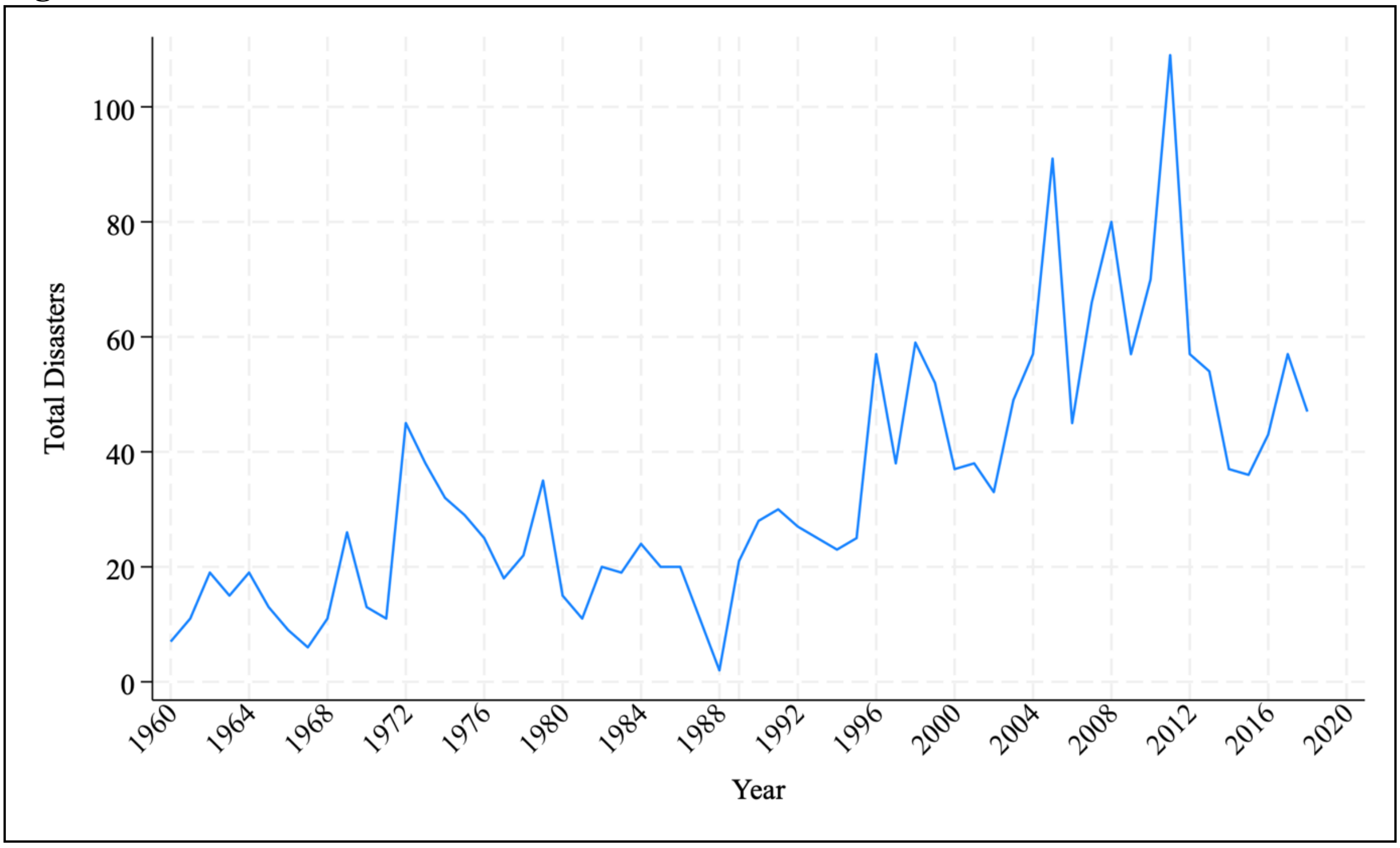

Total disasters measure the annual total number of disasters in the 49 states of our US sample using the FEMA dataset.

**Countries' Appendix**

List of countries in the WVS sample

1. Albania
2. Algeria
3. Argentina
4. Armenia
5. Australia
6. Azerbaijan
7. Bangladesh
8. Belarus
9. Bolivia
10. Bosnia and Herzegovina
11. Brazil
12. Bulgaria
13. Burkina Faso
14. Canada
15. Chile
16. China
17. Colombia
18. Cyprus
19. Czech Republic
20. Dominican Republic
21. Ecuador
22. Egypt
23. El Salvador
24. Estonia
25. Ethiopia
26. Finland
27. France
28. Georgia
29. Germany
30. Ghana
31. Greece
32. Guatemala
33. Haiti
34. Hong Kong
35. Hungary
36. India
37. Indonesia
38. Iran
39. Iraq
40. Italy
41. Japan
42. Jordan
43. Kazakhstan
44. Kuwait
45. Kyrgyzstan
46. Latvia
47. Lebanon
48. Libya
49. Lithuania
50. Malaysia
51. Mali
52. Mexico
53. Moldova
54. Montenegro
55. Morocco
56. Netherlands
57. New Zealand
58. Nigeria
59. North Macedonia
60. Norway
61. Pakistan
62. Palestine
63. Peru
64. Philippines
65. Poland
66. Puerto Rico
67. Romania
68. Russia
69. Rwanda
70. Serbia
71. Slovakia
72. Slovenia
73. South Africa
74. South Korea
75. Spain
76. Sweden
77. Switzerland
78. Taiwan
79. Thailand
80. Trinidad and Tobago
81. Tunisia
82. Turkey
83. Uganda
84. Ukraine
85. United Kingdom
86. United States
87. Uruguay
88. Uzbekistan
89. Venezuela
90. Vietnam
91. Yemen
92. Zambia
93. Zimbabwe

Data Description for WVS

| Variable | Description | Source |
|---|---|---|
| | *Main Variables* | |
| Temperature | Temperature is calculated as the running ten-year standard deviation (or mean) of the temperature reached in each subnational region. | Harris et al. (2020) |
| Satisfied | A measure of well-being based on the WVS question on how satisfied the respondent is with their life. The scale ranges from zero to ten with larger values corresponding to a higher level of satisfaction. | WVS Database (2020) |
| Happy | A measure of well-being based on the WVS question on how happy is the respondent. The answers can range from (0) not "happy at all" to (3) "very happy". | WVS Database (2020) |
| | *Baseline Controls* | |
| Age | The age of the respondent. | WVS Database (2020) |
| Income | The level of income of the respondent. The scale ranges from zero to ten with larger values corresponding to a higher step in the income scale. | WVS Database (2020) |
| Education | The variable measures the highest level of education attained by the respondent. The scale has three categories: low, medium and high (1-3). | WVS Database (2020) |
| Sex | The gender of the respondent. A value of one is assigned for a male and zero for female. | WVS Database (2020) |
| Marital status | Marital status of the respondent. A dummy variable is generated for each of the categories of classification - married, cohabiting, divorced, separated, widowed and single. | WVS Database (2020) |
| Health status | The variable measures the subjective health of the respondent. A dummy variable is generated for each of the categories of classification – very poor, poor, normal, good. | WVS Database (2020) |
| Religiosity | A measure for religiosity based on the WVS question on how important is religion in the respondent's life. The variable is recoded such that the answers range from (0) "not at all important" to (4) "very important". | WVS Database (2020) |
| Number of children | The respondent's total number of children. | WVS Database (2020) |
| Employment status | A dummy variable to indicate the employment status of the respondent. A value of one is assigned if employed, and zero otherwise. | WVS Database (2020) |
| Honesty | A measure for honesty based on the WVS question if it is justifiable to cheat on taxes. The variable is recoded such that the answers range from (0) "always" to (10) "Never". | WVS Database (2020) |
| | *Natural Disasters* | |
| Drought | The total annual occurrence of droughts in each subnational region. | Rosvold and Buhaug (2021) |
| Flood | The total annual occurrence of floods in each subnational region. | Rosvold and Buhaug (2021) |
| Storm | The total annual occurrence of storms in each subnational region. | Rosvold and Buhaug (2021) |
| Total Disasters | The total annual occurrence of droughts, floods, and storms in each subnational region. | Rosvold and Buhaug (2021) |

Data Description for BRFSS

| Variable | Description | Source |
|---|---|---|
| | *Main Variables* | |
| Temperature | Temperature is calculated as the running ten-year standard deviation (or mean) of the temperature reached in each county. | Harris et al. (2020) |
| Life Satisfaction | A measure of well-being based on the BRFSS question "In general, how satisfied are you with your life?". The scale ranges from one to four with larger values corresponding to a higher level of satisfaction. | CDC (2011) |
| | *Baseline Controls* | |
| Age | The age of the respondent. | CDC (2011) |
| Income | The level of income of the respondent. The scale ranges from zero to eight with larger values corresponding to a higher step in the income scale. | CDC (2011) |
| Education | The variable measures the highest level of education attained by the respondent. The scale has six levels with increasing level of education corresponding to a higher value. | CDC (2011) |
| Sex | The gender of the respondent. A dummy variable is generated for males. | CDC (2011) |
| Marital status | Marital status of the respondent. A dummy variable is generated for each of the categories of classification - married, divorced, separated, and widowed. | CDC (2011) |
| Health status | The variable measures the subjective health of the respondent. A dummy variable is generated for each of the categories of classification – poor, fair, very good, excellent. | CDC (2011) |
| Number of children | The respondent's total number of children. | CDC (2011) |
| Employment status | A dummy variable to indicate the employment status of the respondent. A value of one is assigned if employed, and zero otherwise. | CDC (2011) |
| | *Natural Disasters* | |
| Drought | The total annual occurrence of droughts in each county. | Rosvold and Buhaug (2021) |
| Flood | The total annual occurrence of floods in each county. | Rosvold and Buhaug (2021) |
| Storm | The total annual occurrence of storms in each county. | Rosvold and Buhaug (2021) |
| Total Disasters | The total annual occurrence of droughts, floods, and storms in each county. | Rosvold and Buhaug (2021) |

**References for Data**